\documentclass[aps,prd,superscriptaddress,12pt,showpacs,notitlepage]{revtex4}
\usepackage{amsmath,amssymb,mathrsfs}
\usepackage{graphicx}

\newcommand{\MeV}{\textrm{MeV}}

\newcommand{\be}{\begin{equation}}
\newcommand{\ee}{\end{equation}}
\newcommand{\bea}{\begin{eqnarray}}
\newcommand{\eea}{\end{eqnarray}}
\newcommand{\pa}{\partial}

\begin{document}
\title{BPS Skyrme model and nuclear binding energies}
\author{C. Adam}
\affiliation{Departamento de F\'isica de Part\'iculas, Universidad de Santiago de Compostela and Instituto Galego de F\'isica de Altas Enerxias (IGFAE) E-15782 Santiago de Compostela, Spain}
\author{C. Naya}
\affiliation{Departamento de F\'isica de Part\'iculas, Universidad de Santiago de Compostela and Instituto Galego de F\'isica de Altas Enerxias (IGFAE) E-15782 Santiago de Compostela, Spain}
\author{J. Sanchez-Guillen}
\affiliation{Departamento de F\'isica de Part\'iculas, Universidad de Santiago de Compostela and Instituto Galego de F\'isica de Altas Enerxias (IGFAE) E-15782 Santiago de Compostela, Spain}
\author{A. Wereszczynski}
\affiliation{Institute of Physics,  Jagiellonian University,
Reymonta 4, Krak\'{o}w, Poland}


\begin{abstract}
We use the classical BPS soliton solutions of the BPS Skyrme model together with corrections from the collective coordinate quantization of spin and isospin, the electrostatic Coulomb energies, and a small explicit breaking of the isospin symmetry - accounting for the proton-neutron mass difference - to calculate nuclear binding energies. We find that the resulting binding energies are already in excellent agreement with their physical values for heavier nuclei, demonstrating thereby that the BPS Skyrme model is a distinguished starting point for a detailed quantitative investigation of nuclear and low-energy strong interaction physics.
\end{abstract}

\maketitle

{\bf Introduction:} 
The Skyrme model \cite{skyrme}, \cite{manton} is a nonlinear field theory meant to describe the low-energy limit of strong interaction physics. Its primary fields are mesons, whereas baryons appear as solitons (skyrmions), classified by an integer-valued topological degree which is identified with baryon number \cite{thooft}. In the two flavor case, the field of the Skyrme model takes values in SU(2), and one natural area of applications is nuclear physics, providing a possible basis for a unified field theoretic description of nuclei. And indeed, shortly after the analysis of the nucleons in terms of the simplest skyrmion (hedgehog) \cite{nappi}, the Skyrme model was employed to describe the deuteron \cite{braaten} and some additional light nuclei \cite{carson}. More recently, nuclear excitation spectra have been studied within the Skyrme model, e.g., in \cite{wood}, with reasonable success. One common problem in applying the standard Skyrme model to nuclei are the resulting too large binding energies. There exists a lower energy bound linear in the baryon number $B$ (Faddeev-Bogomolnyi bound, see, e.g., \cite{rybakov-book}, \cite{manton-book}), but skyrmions do not saturate it. The simplest $B=1$ skyrmion is about 23\% above the bound, whereas for higher $B$ this deviation is lowered, to less than 4\% in the limit of large $B$. So, the binding energies per baryon number of higher skyrmions are on the level of 10\%, in striking contrast to the small values for physical nuclei. Also, their calculation requires demanding numerics, in contrast to the almost entirely analytical results we shall present below.

Recently, some of us found a certain Skyrme submodel ("BPS Skyrme model") which has both a Bogomolnyi-type energy bound and infinitely many solutions solving a first-order (BPS) equation and saturating the bound \cite{BPS-Sk}.
This observation leads to the natural proposal \cite{BPS-Sk}, \cite{marleau1} to treat the solitons of the BPS submodel as the leading contributions to nuclear masses, further supported by the observation that the submodel has the symmetries of an incompressible ideal liquid \cite{BPS-Sk}, \cite{fosco}, serving as a field theory realization of the liquid drop model of nuclei. Relatively small corrections to the nuclear masses (and, therefore, small but nonzero binding energies) may then be described by some further contributions. It is the purpose of the present letter to determine the resulting nuclear binding energies. Concretely, we shall include the effects of the collective coordinate quantization of spin and isospin, the Coulomb energy and an explicit small breaking of the isospin symmetry, all of which are completely natural contributions within Skyrme theory. We will find that the resulting nuclear binding energies are already in very good agreement with the experimental values for heavy nuclei, demonstrating that the BPS Skyrme model together with standard Skyrme technology provide an excellent starting point for a detailed study of nuclear physics and, more generally, strong interaction physics at low energies (see also \cite{vibr}), both at a conceptual and a calculational level.

{\bf BPS Skyrme model:}
The standard Skyrme model lagrangian consists of two terms, a term quadratic in first derivatives (the sigma model term ${\cal L}_2$), and the Skyrme term ${\cal L}_4$ quartic in first derivatives, needed to balance the scaling instability such that soliton solutions may exist. The effective field theory (EFT) philosophy allows (and requires), in principle, to add all possible terms compatible with the basic symmetries, and that is precisely what is done in a perturbative approach (chiral perturbation theory). But for a nonperturbative framework like the Skyrme model, some selection principle is required highlighting the relevant physics.  Here we shall require that the lagrangian is no more than quadratic in time derivatives such that a standard Hamiltonian exists. Together with the obvious requirement of Poincare invariance, this severely restricts possible terms in the lagrangian. Essentially, only a potential term ${\cal L}_0$ and a certain term ${\cal L}_6$ sextic in derivatives \cite{sextic} may be added to the two standard terms ${\cal L}_2$ and ${\cal L}_4$. So we consider the lagrangian
${\cal L}={\cal L}_0 + {\cal L}_2 + {\cal L}_4 + {\cal L}_6$
(our Minkowski metric conventions are ${\rm diag} (g_{\mu\nu}) = (+,-,-,-)$)
where 
\be
{\cal L}_2 = \lambda_2 {\rm tr}\; \pa_\mu U \pa^\mu U^\dagger \equiv - \lambda_2 {\rm tr}\; L_\mu L^\mu \; ,
\quad {\cal L}_4 = \lambda_4 {\rm tr}\; \left( [L_\mu ,L_\nu ] \right)^2
\ee
and 
\be
{\cal L}_0 = -\lambda_0 V({\rm tr}\; U) \; , \quad {\cal L}_6 = -\lambda_6 \left( \epsilon^{\mu\nu\rho\sigma} {\rm tr}\; L_\nu L_\rho L_\sigma \right)^2 
\equiv -(24 \pi^2)^2 \lambda_6 {\cal B}_\mu {\cal B}^\mu .
\ee
Here $U: {\mathbb R}^3 \times {\mathbb R} \to {\rm SU(2)}$ is the Skyrme field and $L_\mu = U^\dagger \pa_\mu U$ the left-invariant Maurer-Cartan current. Further, the $\lambda_n$ are non-negative coupling constants, and ${\cal B}_\mu$ is the topological or baryon number current giving rise to the topological degree (baryon number) $B\in {\mathbb Z}$,
\be
{\cal B}^\mu = \frac{1}{24 \pi^2 } \epsilon^{\mu\nu\rho\sigma} {\rm tr}\; L_\nu L_\rho L_\sigma \; ,\quad B= \int d^3 x {\cal B}^0 .
\ee
 The terms ${\cal L}_2$, ${\cal L}_4$ and ${\cal L}_6$ are invariant under the chiral transformations $U \to WUW'$, $W,W' \in $ SU(2), whereas the potential breaks this symmetry down to the diagonal (isospin) subgroup $U\to WUW^\dagger$. Further, we assume from now on that the potential $V({\rm tr} \; U)$ is non-negative and has one unique vacuum at $U={\bf 1}$, i.e., $V(U={\bf 1})=0$. 

The BPS Skyrme model is the limit $\lambda_2 = \lambda_4 =0$ of the above theory, with lagrangian (conveniently defining $\lambda_6 = \lambda^2 /(24)^2$ and $\lambda_0 = \mu^2$) 
\be \label{submodel} 
{\cal L}_{06} = - \pi^4 \lambda^2  {\cal B}_\mu {\cal B}^\mu - \mu^2 V({\rm tr}\; U) .
\ee
The static energy functional of this theory,
\be \label{BPS-en-funct}
E = \int d^3 x \left( \pi^4 \lambda^2  {\cal B}_0^2 + \mu^2 V({\rm tr}\; U) \right)
\ee
 has both a Bogomolnyi bound 
and infinitely many BPS solutions saturating the bound \cite{BPS-Sk}, \cite{speight}.
This crucial observation, together with the small binding energies of physical nuclei, leads to the plausible assumption that the Skyrme model relevant for nuclear physics belongs to a region in EFT parameter space where the terms 
${\cal L}_0$ and ${\cal L}_6$ give the main contributions to nuclear masses, whereas the remaining terms provide rather small corrections. More generally, the two terms ${\cal L}_0$ and ${\cal L}_6$ could be the most important ones in regions of not too small baryon and energy densities.
As a consequence, the submodel (\ref{submodel}) should provide a good starting point (lowest order approximation) and give already reasonable results for certain static properties of nuclei. It is the purpose of the present letter to further investigate this proposal.
Clearly, the submodel (\ref{submodel}) cannot provide a detailed description of all strong interaction physics, there is, e.g., no perturbative pion dynamics (absence of ${\cal L}_2$). More generally, in near vacuum regions the hierarchy of terms will be reversed (i.e., the term ${\cal L}_2$ will always be dominant near the vacuum, even for parameter choices such that ${\cal L}_6$ is dominant for high baryon density). 
Now we have to make several choices. First of all, for simplicity we choose the standard Skyrme potential
\be \label{Sk-pot}
V = \frac{1}{2} {\rm tr} \; ({\bf 1}-U) = 1-\cos \xi 
\ee
(where
$
U= \cos \xi +i\sin \xi \; \vec n \cdot \vec \tau$, $\vec n^2 =1,
$
and $\vec \tau$ are the Pauli matrices),
although without the term ${\cal L}_2$ there is no direct relation to the pion masses, and other choices are possible. 
Secondly, we have to choose the shapes (symmetries) of our skyrmion solutions, because
due to the many symmetries of the energy functional (\ref{BPS-en-funct}), there exist solutions with different shapes \cite{fosco}.
For simplicity, we shall choose the axially symmetric ansatz
\begin{equation} \label{ansatz}
  \xi = \xi (r) \; , \quad \vec n = (\sin \theta \cos n\phi ,\sin \theta \sin n\phi , \cos \theta )
\end{equation}
where $(r,\theta, \phi)$ are spherical polar coordinates. Here our point of view is that, although most nuclei probably do not have this symmetric shape and their determination requires the solution of a much more involved variational problem, the symmetric ansatz (\ref{ansatz}) may still give a reasonable approximation provided that the deviation of physical nuclei from a symmetric shape is no too pronounced.
The ansatz (\ref{ansatz}) leads to $B=n$ and to the first order ODE for $\xi (r)$
\begin{equation}
 \frac{n\lambda}{2\mu}  \frac{\sin^2 \xi}{\sqrt{V}} \frac{d\xi}{dr}  =  - r^2 .
\end{equation}
Specifically, for the standard Skyrme potential (\ref{Sk-pot}) the solution is compact and given by a simple arcsine function, see \cite{BPS-Sk}. We remark that the simple analytic form of the solutions permits the analytic calculation of further physical quantities like form factors or charge radii for any baryon charge \cite{BPS-Sk}.

{\bf Nuclear binding energies:}
For the static energy (mass) $E$ of a nucleus, we shall take into account the following contributions,
$
E=E_{\rm sol} + E_{\rm rot} + E_{\rm C} + E_{\rm I}
$,
where $E_{\rm sol}$ is the classical soliton energy, which for the standard Skyrme potential is \cite{BPS-Sk}
\begin{equation}
E_{\rm sol} = \frac{64 \sqrt 2 \pi}{15} \mu \lambda |n|.
\end{equation}
 $E_{\rm rot}$ comes from the collective coordinate quantization of rotations and iso-rotations, giving rise to spin and isospin of the nuclei. $E_{\rm C}$ is the electrostatic Coulomb energy of the nucleus, whereas $E_{\rm I}$ is a contribution from a small, explicit isospin breaking, taking into account the mass difference between proton and neutron.

{\bf Spin and isospin:}
The (semi-classical) quantization of rotations and iso-rotations is obviously required for a consistent description of nuclei, because both spin and isospin are relevant quantum numbers of nuclei.  As usual, the rigid rotor quantization of spin and isospin proceeds by introducing the rotational and iso-rotational degrees of freedom about a static soliton via
$
U(t,\vec x)= {\rm A}(t)U_0(R_{\rm B}(t) \vec x){\rm A}^\dagger (t)
$
(where ${\rm A, B} \in $ SU(2), $R_{\rm B} \in$ SO(3), $U_0$ $\ldots$ soliton), and by inserting this expression into the Skyrme lagrangian. The variables parametrizing $A$ and $B$ and their canonical momenta are then quantized via canonical quantization.
The result is the standard quantized rigid rotor hamiltonian both for spin and for isospin, and different nuclei are identified with different eigenstates of the rigid rotor, i.e., 
\be \label{nuc-state}
|X\rangle = |jj_3 l_3\rangle |ii_3k_3\rangle
\ee
where $X$ is a nucleus, $\vec J $ ($\vec L$) is the space-fixed (body-fixed) angular momentum,
$\vec I$ ($\vec K$) is the space-fixed (body-fixed) isospin angular momentum, and $j, j_3, l,l_3$ and $i ,i_3 ,k, k_3$ are the corresponding eigenvalues. Finally, $|jj_3l_3\rangle$ and $|ii_3k_3\rangle$ are the eigenstates of the rigid rotor (Wigner D functions) for spin and isospion, respectively.
For soliton solutions $U_0$ with some symmetries, certain combinations of the transformations $A$ and $B$ will act trivially on $U_0$, and the corresponding combinations of collective coordinates are absent from the quantum mechanical rigid rotor Hamiltonian. Further, these transformations should act trivially, up to a phase, on the nuclei described by the soliton $U_0$ (Finkelstein-Rubinstein constraints; a detailed discussion for the standard Skyrme model may be found in \cite{krusch}). 

For the ansatz (\ref{ansatz}) used here, the corresponding moments of inertia are well-known \cite{nappi}, \cite{braaten}, \cite{hou-mag}.
Concretely, for $n=B=1$, the resulting skyrmion (hedgehog) is spherically symmetric, i.e., an arbitrary rotation can be undone by an iso-rotation (and vice versa). As a consequence, only three of the six collective coordinates (e.g., only spin) appear. Further, the body-fixed moments of inertia tensor is proportional to the identity,
\begin{equation} \label{Jota-sym}
{\cal J}_{ij} =\delta_{ij} {\cal J}, \; \; {\cal J} = \frac{4\pi}{3} \lambda^2 \int dr \sin^4 \xi \, \xi_r^2 = \frac{2^8 \sqrt{2} \pi}{15\cdot 7} \lambda \mu \left( \frac{\lambda}{\mu}\right)^\frac{2}{3}
\end{equation} 
and the resulting quantum mechanical hamiltonian of a spherical top leads to the energy
\begin{equation}
E _{\rm rot}= \frac{1}{2{\cal J}} \hbar^2 j(j+1) .
\end{equation}

For $n=B> 1$, the quantum mechanical hamiltonian essentially consists of
two copies (spin and isospin) of a symmetric top (rigid rotor with ${\cal J}_{ij} = {\cal J}_i \delta_{ij}$, and generically ${\cal J}_1 = {\cal J}_2 \not= {\cal J}_3$). 
The axial symmetry implies that a rotation about the three-axis (by an angle $\varphi$) can be undone by an isospin rotation (by an angle $n\varphi$), so the corresponding generator ($L_3$ or $K_3$) should be taken into account only once. The resulting energy is
\begin{equation} \label{E-rot}
E_{\rm rot} = \frac{\hbar^2}{2} \Bigl( \frac{j (j+1)}{{\cal J}_{1}} + \frac{i (i+1)}{{\cal I}_{1}} 
+ \bigl( \frac{1}{{\cal I}_{3}} - \frac{1}{{\cal I}_{1}} - \frac{n^2}{{\cal J}_{1}} \bigr) k_3^2 \Bigr) 
\end{equation}
where ${\cal I}_{ij} = {\cal I}_i \delta_{ij}, {\cal I}_1 = {\cal I}_2 \not= {\cal I}_3$ is the isospin moments of inertia tensor, and
\bea
&& {\cal I}_3 = \frac{4\pi}{3} \lambda^2 \int dr \sin^4 \xi \, \xi_r^2 = |n|^{-\frac{1}{3}}{\cal J} \\
&&  {\cal I}_1 = \frac{3n^2 + 1}{4} {\cal I}_3 \; , \quad  {\cal J}_1 = {\cal J}_3 = n^2 {\cal I}_3 
\eea
and ${\cal J}$ is defined in (\ref{Jota-sym}). Further, the axial symmetry implies
\be
(L_3 + n K_3) |X\rangle =0 \quad \Rightarrow \quad  l_3 + n k_3 =0.
\ee
This constraint, together with $j\ge |l_3|$, leads to unacceptably large angular momenta for physical nuclei for $k_3 \not= 0$. We, therefore, assume $k_3 =0$ in what follows, so the axially symmetric ansatz cannot be used for nuclei with odd baryon number $B$, because such nuclei are fermions with half-odd integer values of $k_3$. We, therefore, restrict our discussion to nuclei with even $B=n$ (this is just a technical restriction resulting from our ansatz (\ref{ansatz}); nuclei with odd $B>1$ have to be described by solitons with other symmetries). Our final results for the (iso-) rotational excitation energies are, for $B=n=1$ (where $j=\frac{1}{2}$),
\begin{equation}
E_{\rm rot} =  \frac{105}{512 \sqrt 2 \pi} \frac{3}{4} \frac{\hbar^2}{\lambda^2 \left( \frac{\mu}{\lambda} \right)^{1/3}}
\end{equation}
and for $B=n>1$ (where, as in \cite{marleau}, we assume that the lowest excitation energy for a given $i_3$ obeys $i=|i_3|$)
\be
E_{\rm rot} = \frac{105}{512 \sqrt 2 \pi} \frac{\hbar^2}{\lambda^2 \big( \frac{\mu}{\lambda n} \big)^{1/3}} \Bigl( \frac{ j (j+1)}{n^2} +\frac{4|i_3|(|i_3|+1)}{3n^2 +1} \Bigr) .
\ee

{\bf Coulomb energies:}
The Coulomb energy contributions to the binding energies are especially relevant for heavier nuclei. The Coulomb energy is
\begin{equation}
 E_{\rm C}=\frac{1}{2 \varepsilon_0} \int d^3 x d^3 x' \frac{\rho(\vec r) \rho(\vec r\,')}{4 \pi|\vec r - \vec r\,'|}.
\end{equation}
where $\rho$ is the electric charge density of the nucleus, i.e., the expectation value w.r.t. nuclear wave functions $| X \rangle$ of the electric charge density operator \cite{cal-wit}
\be
\hat{\rho} =\frac{1}{2} {\cal B}^0 + {\mathbb J}_3^0 
\ee
where 
\begin{equation} \label{B0}
{\cal B}^0 = - \frac{n}{2 \pi^2 r^2} \sin^2 \xi \, \xi_r ,
\end{equation}
 is the topological charge density and $ {\mathbb J}_3^0 $ is the time component of the third isospin current density operator ${\mathbb J}^\mu_3$. Again, a collective coordinate quantization has been performed.
For the second term ${\mathbb J}^0_3 = {\mathbb J}^0_3 (\vec \Omega, \vec \omega, \vec R)$ a complication arises because it
depends both on the spin and isospin angular velocities $\vec \omega$ and $\vec \Omega$ and on the isospin collective coordinates $\vec R$, so a Weyl ordering is required (see \cite{Ding} for details). Indeed, 
a long but straight-forward calculation leads to the explicitly Weyl-ordered expression
\begin{equation} \label{iso-dens-op}
{\mathbb J}_3^0 = - \frac{\lambda^2}{4r^2} \xi_r^2 \sin^4 \xi \bigl( (R_i \Omega_j + \Omega_j R_i)  \,  {\cal A}_{ij}  + (R_i \omega_j + \omega_j R_i)  \, {\cal B}_{ij} \bigr) ,
\end{equation} 
where ${\rm A}^\dagger \tau_3 {\rm A} = R_i \tau_i $.
Further, ${\cal A}_{ij}$ and ${\cal B}_{ij}$ are matrices depending on the angles $\theta $ and $\phi$ of the spherical polar coordinates, which we do not display here. The presence of both collective coordinates $R_i$ and angular velocities (related to the body-fixed angular momenta $\vec K$ and $\vec L$) is very important here because of the identity $I_3 = - \vec R\cdot \vec K$, so only the presence of both $\vec R$ and $ \vec \Omega$ in (\ref{iso-dens-op}) allows to recover the correct (space-fixed) isospin dependence for nuclear matrix elements $\langle X | {\mathbb J}^0_3 | X\rangle$. We differ in this respect from Ref. \cite{marleau}, because their isospin density operator only depends on the angular velocities. The final result for the matrix elements is (details of the calculation can be found in \cite{bound-PRC})
\be \label{matrix-el}
\langle X |{\mathbb J}_3^0 | X \rangle 
= \frac{\lambda^2}{2r^2} \xi_r^2 \sin^4 \xi \frac{1}{{\cal I}_3} \left( \frac{2(n^2 + \cos^2 \theta )}{3n^2 +1}i_3 - \frac{(n^2 +1)(1-3\cos^2 \theta )}{3n^2 +1} \langle X|R_3 K_3 |X\rangle \right) .
\ee
It may be checked easily that the contribution to the electric charge is just $i_3$, i.e., $\int d^3 x \langle X| {\mathbb J}_3^0 |X\rangle =i_3$, such that the electric charge is $Z = \int d^3 x \langle X|\hat{\rho} |X\rangle = (B/2) + i_3$, as it must be (observe that the r.h.s. of (\ref{matrix-el}) depends on $\lambda$ both explicitly and implicitly via the profile function $\xi$, such that this $\lambda$ dependence cancels in the integral (the charge)). The matrix element $\langle X|R_3 K_3 |X\rangle $ does not contribute even if $k_3 \not= 0$, because its prefactor integrates to zero.  
For $n=B>1$, we may ignore this term even for the electric charge densities, because our nuclear wave functions obey $K_3 |X\rangle =0$.

For the calculation of the Coulomb energy one now has to perform the usual multipole expansion of the Coulomb potential
and expand the charge density into spherical harmonics, see, e.g.,  \cite{marleau}.
The result is,  for $n=B=1$,
\begin{eqnarray}
\mbox{proton:} \quad &E_C^{\rm p} =& 
\frac{1}{\sqrt{2} \pi\varepsilon_0} \Bigg( \frac{\mu}{\lambda } \Bigg)^{1/3} \Bigg(
\frac{128}{315 \pi^2}  + \frac{156625}{1317888} \Bigg) 
 \\
\mbox{neutron:} \quad &E_C^{\rm n} =&  
\frac{1}{\sqrt{2} \pi\varepsilon_0} \Bigg( \frac{\mu}{\lambda } \Bigg)^{1/3} \Bigg(
\frac{128}{315 \pi^2}  - \frac{53585}{1317888} \Bigg)
\end{eqnarray} 
and for $n=B>1$
\begin{equation}
E_{\rm C}= \frac{1}{\sqrt{2} \pi\varepsilon_0} \Bigg( \frac{\mu}{\lambda n} \Bigg)^{1/3}
\Bigg( 
\frac{128}{315 \pi^2 }  n^2 + \frac{245}{1536 }  n \; i_3 +  \frac{805}{5148 }  i_3^2 
 + \frac{7}{429 }  \frac{i_3^2}{(1+3n^2)^2} \Bigg) .
\end{equation}

{\bf Isospin breaking:}
The mass difference between neutron and proton may be traced back, at a microscopic level, to the mass difference between up and down quark, but within the Skyrme model it should result from isospin-breaking terms in the effective Skyrme Lagrangian \cite{iso-breaking}. 
A detailed discussion of this issue is beyond the scope of the present article. Here we just take into account the leading order effect of the isospin breaking, which is obviously given by the  hamiltonian ${\cal H}_{\rm I} = a_{\rm I} I_3$ with the resulting energy 
$E_{\rm I} = a_{\rm I} i_3$  where  $a_{\rm I}<0$.
In principle, it should be possible to calculate the constant $a_{\rm I}$ from the microscopic theory, but here we shall treat it as a free parameter. 

{\bf Results and discussion:}
For an explicit calculation of nuclear masses, 
 we have to determine numerical values for the three parameters $\lambda $, $\mu$ and $a_{\rm I}$ of our model. Concretely, we fit to the nuclear masses of the proton, neutron, and the nucleus with magical numbers $^{138}_{\hspace{0.14cm}56}{\rm Ba}$, with masses
$M_{\rm p} =938.272 \; \MeV  $, $M_{\rm n}  = 939.565 \; \MeV  $, and
$M(^{138}_{\hspace{0.14cm}56}{\rm Ba})  = 137.894 \; \textrm u$ where  $\textrm{u} = 931.494 \; \MeV $.
With
$
\hbar = 197.327 \; \textrm{MeV  fm}  $,
$\varepsilon_0 = e^{-1} \; 8.8542 \cdot 10^{-21} \textrm{MeV}^{-1}  \textrm{fm}^{-1}$, and
$e = 1.60218 \cdot 10^{-19}$,
this leads to the parameter values
\begin{equation}
\lambda \mu =   48.9902 \; \MeV , \;  \left( \frac{\mu}{\lambda} \right)^\frac{1}{3} =  
0.604327 \; \textrm{fm}^{-1}, \;
a_{\rm I} = -1.68593 \; \MeV .
\end{equation}
With these values, we now may determine the masses (energies $E_X$) of many more nuclei,
$
E_X = E_{\rm sol} + E_{\rm rot} + E_{\rm C} + E_{\rm I} 
$
where, however, we prefer to determine (and plot), instead, the nuclear binding energies
\begin{equation}
E_{{\rm B},X} = Z E_{\rm p} + N E_{\rm n} - E_X .
\end{equation} 
Here, 
$ i_3 = \frac{1}{2}(Z-N) $, $ n=B\equiv A = Z+N$ ($A\ldots$ nuclear weight number, $Z\ldots$ number of protons, $N\ldots$ number of neutrons), and $E_{{\rm B},X}(A,Z,j)$ may be expressed as a function of the 
nuclear charge $Z$, atomic weight number $A$ and spin $j$.

\begin{figure}[h]
\begin{center}
\includegraphics[width=0.6\textwidth]{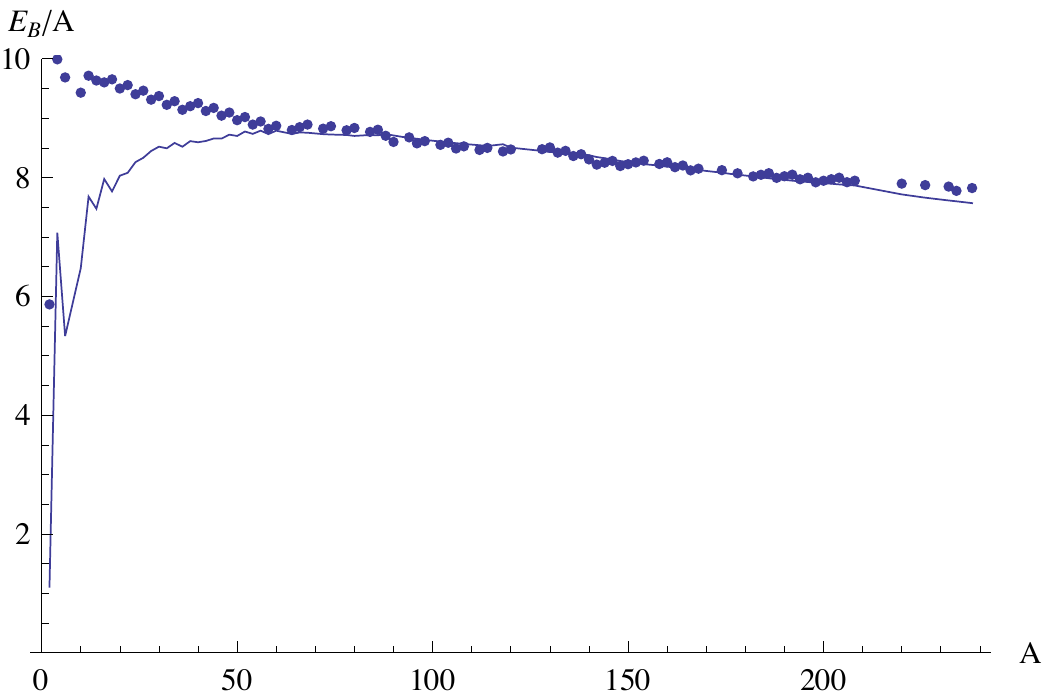}
\caption{Binding energies per nucleon in MeV. The experimental values are given by the solid line whereas the dots represent the results from our model.}
\label{FigEB}
\end{center}
\end{figure}

For a comparison with experimental values we now follow the strategy of \cite{marleau}. I.e., for each fixed value of the atomic weight number $A$, we choose the values of $Z$ and $j$ corresponding to the most abundant nucleus. For the resulting nuclei we then compare the binding energies per atomic weight number, $E_{\rm B} / A$, with their experimental values \cite{Masses}, see Fig. \ref{FigEB}.

We find that our simple model already provides an excellent description of nuclear binding energies for higher atomic weight numbers, whereas it overestimates the binding energies of small nuclei. This is to be expected, because the BPS Skyrme model is based on two terms in the effective lagrangian which are related to {\em collective} phenomena (the topological current density squared and the chiral-symmetry breaking potential) and has the symmetries of an incompressible ideal liquid, relating it to the nuclear liquid drop model, again a collective description of nuclei. Further, our model has just three fit parameters, all of which follow from the same unified field theoretic description. For small nuclei, single-particle properties or pion degrees of freedom will be more important and require an extension of the model. These extensions are, however, completely natural within the field theoretic framework of (generalized) Skyrme theory. Indeed, our proposal entirely belongs to the EFT framework, just assuming that there exists a privileged region in EFT parameter space which enhances the role of the terms ${\cal L}_6$ and ${\cal L}_0$, and investigating the consequences of this assumption. 

Similar calculations using our BPS model have been done in \cite{marleau} with other premises (e.g., using different potentials) leading to different results (a detailed discussion will be given elsewhere). First of all, their expressions for the isospin current density operator and, consequently, their Coulomb energies, differ from ours. We believe, in fact, that their isospin current density operator is not adequate. Secondly, they include contributions from the terms ${\cal L}_2$ and ${\cal L}_4$ to the soliton energies in a perturbative way. But  the axially symmetric ansatz (\ref{ansatz}) leads to perturbative contributions to the classical soliton energies which grow much faster than linear with the baryon number. This fast growth implies a strong influence on the nuclear binding energies and, consequently, unphysically small values of their coupling constants. A correct treatment, which will be the result of a more involved variational problem, should lead to contributions which are linear in the bayon charge as befits contributions to classical soliton energies, and which therefore do not contribute to the binding energies in a first order perturbation. This is one of the reasons why we did not consider them in this letter. 

To summarize, we have found ample evidence for the claim that the BPS Skyrme model provides an excellent point of departure for a renewed effort to apply Skyrme theory to the investigation of nuclear and strong interaction physics, which should complement existing approaches and lead to a significant improvement of quantitative results in many situations where Skyrme models have already been employed with success.      
We think we have presented an important step towards the ultimate goal of Skyrme theory of providing a unified field theoretic description of a vast area of phenomena of nuclear and hadronic physics. 

{\bf Acknowledgement:}
The authors acknowledge financial support from the Ministry of Education, Culture and Sports, Spain (grant FPA2008-01177), 
the Xunta de Galicia (grant INCITE09.296.035PR and
Conselleria de Educacion), the
Spanish Consolider-Ingenio 2010 Programme CPAN (CSD2007-00042), and FEDER. 
CN thanks the Spanish
Ministery of
Education, Culture and Sports for financial support (grant FPU AP2010-5772).
Further, AW was supported by polish NCN (National Science Center) grant DEC-2011/01/B/ST2/00464 (2012-2014). The authors thank M. Speight and N. Manton for helpful discussions.

\end{document}